\documentstyle[aps,preprint,epsfig]{revtex}

\begin{document}
\preprint{TU-598}
\title{No-Scale Scenario with Non-Universal Gaugino Masses}
\author{Shinji Komine\footnote{komine@tuhep.phys.tohoku.ac.jp}
and Masahiro Yamaguchi\footnote{yama@hep.phys.tohoku.ac.jp}}
\address{Department of Physics, Tohoku University,
Sendai 980-8578, Japan}
\date{July, 2000}
\maketitle
\begin{abstract}

Phenomenological issues of no-scale structure of K\"{a}hler
potential are re-examined, which arises in various approaches to
supersymmetry breaking. When no-scale boundary conditions are given
at the Grand Unified scale and universal gaugino masses are
postulated, a bino mass is quite degenerate with right-handed
slepton masses and the requirement that the lightest superparticle
(LSP) be neutral supplemented with slepton searches at LEP200 severely
constrains allowed mass regions of superparticles.  The situation
drastically changes if one moderately relaxes the assumption of the
universal gaugino masses. After reviewing some interesting scenarios
where non-universal gaugino masses arise, we show that the
non-universality diminishes the otherwise severe constraint on the
superparticle masses, and leads a variety of superparticle mass
spectra: in particular the LSP can be a wino-like neutralino, a
higgsino-like neutralino, or even a sneutrino, and also left-handed 
sleptons can be lighter than right-handed ones.

\end{abstract} 
\clearpage

\section{Introduction}

One of the most important phenomenological issues in supersymmetric
(SUSY) Standard Models (SSMs) is to identify the mechanisms of
supersymmetry breaking in the hidden sector and its mediation to the
SSM sector (observable sector). Soft supersymmetry breaking
masses which arise in effective theories after integrating over the
hidden sector are in fact constrained from various requirements. 
For instance, they
should lie in the range of $10^2$-$10^3$ GeV to solve the naturalness
problem in the Higgs sector which is responsible for the electroweak
symmetry breaking, and satisfy mass bounds given by collider
experiments. They should also satisfy flavor-changing-neutral-current 
(FCNC) constraints as
well. Furthermore, if the lightest superparticle (LSP) is stable,
which is often the case, cosmological arguments require it be
electrically neutral and $SU(3)_c$ singlet.

The structure of the soft scalar masses is characterized by
the K\"{a}hler potential.  In this paper, we shall focus on a special
class of the K\"{a}hler structure in which the hidden sector and the
observable sector are separated from each other in the K\"{a}hler 
potential $K$ as follows:
\begin{equation}
e^{-K/3} =f_{hid}(z,z^*) +f_{obs}(\phi, \phi^*), \label{eq:no-scale-structure}
\end{equation}
where $z$ and $\phi$ symbolically represent fields in the hidden and
observable sector, respectively. The first example which exhibits 
this form of the K\"ahler potential is a so-called {\em no-scale} 
model\cite{no-scale-model},
and thus we call it the {\em no-scale} structure.  
The characteristics of the K\"ahler potential in the no-scale form 
is that the soft 
SUSY breaking scalar masses vanish (as the vacuum energy vanishes) 
and gaugino masses are a dominant
source of SUSY breaking mass. Of course, this mass pattern is given at the 
energy scale where the soft masses are given, and the renormalization 
group effects due to the non-vanishing gaugino masses raise
the masses of the scalar superparticles at the weak scale.

The no-scale structure of the K\"ahler potential is obtained in many types
of models. As we will see in the next section, such models include the 
(tree-level) K\"ahler potential of  simple Calabi-Yau compactification of 
the heterotic string theories \cite{Witten} both in
the weak- and strong-coupling regimes, the 
splitting Ansatz of the hidden 
and observable sectors in the superspace density in a supergravity 
formalism \cite{IKYY},
and the geometrical splitting of the two sectors in a brane scenario 
\cite{Randall-Sundrum,Luty-Sundrum}.

In this paper we revisit some phenomenological issues of the models
with the no-scale boundary conditions. This class of models has
closely been investigated in the literature. A particular attention was paid
to the {\em minimal} case
where the boundary conditions are given at the Grand Unified Theory (GUT) 
scale of 
$2 \times 10^{16}$ GeV and the gaugino masses are assumed to be universal at 
this energy
scale. In this case  the mass spectrum of superparticles
is very constrained, and the bino mass is almost degenerated with those of
the right-handed sleptons. In fact it was shown that the neutralino can 
be the LSP only when its mass is less than about 120 GeV 
\cite{IKYY,KLNPY,Drees-Nojiri}: otherwise the
stau would be the LSP which is charged, and thus not allowed if it is stable.
We will revise this result, emphasizing that the present experimental 
bounds already exclude the large $\tan \beta$ case, leaving only $\tan \beta 
\lesssim 8$.

One of the main points in this paper is that slight modifications of
the minimal scenario will drastically change mass spectrum of the
superparticles.  In particular, we shall devote ourselves to the case
where the gaugino masses are non-universal at the GUT scale. We will
first review several cases that the non-universality of the gaugino
masses result.  Then we will discuss its phenomenological
implications. Most remarkably relaxing the universality condition
within a factor of two or so will result in a variety of mass spectra.
In particular the LSP can be not only the bino-dominant neutralino,
but also a wino or higgsino-dominant neutralino, or an admixture of
the gaugino and the higgsino, or even a sneutrino. Furthermore the
severe upperbound on the masses of the superparticles no longer
exists. Thus we expect superparticle phenomenology in this case is
much richer than the minimal case.

The paper is organized as follows.
In the subsequent section, we review some examples which possess the
no-scale K\"ahler potential. In section 3, we re-examine the case where
the no-scale boundary conditions are given at the GUT scale and gaugino
masses are universal at the scale, and show that the superparticle mass
spectrum is very restrictive and tight constraints already exclude much 
of the parameter space. In section 4, we argue that the very constrained
mass spectrum can be relaxed by several ways, and then we focus on one
of them, namely the case with non-universal gaugino masses. After recalling
some mechanisms to realize the non-universality of the gaugino masses, we
consider its phenomenological implications. The final section is
devoted to conclusions. 

\section{No-Scale Boundary Conditions}

In this section, we would like to review some models which have the no-scale
K\"ahler potential. The first model is the no-scale 
model \cite{no-scale-model} with the K\"ahler
potential 
\begin{equation}
  K=-3\ln(T+T^*-\phi^*\phi)
\label{eq:no-scale-model}
\end{equation}
and the superpotential
\begin{equation}
  W=W(\phi),
\end{equation}  
where $T$ is a hidden sector field responsible for the SUSY
breaking and $\phi$ is a generic matter field. Here and in the 
following, we use a unit that the
reduced Planck scale $M_{pl}=2.4\times 10^{18}$ GeV set to unity.
With the above K\"ahler potential and superpotential, one can compute
the scalar potential in supergravity and find that 
\begin{equation}
  V=\frac{1}{3(T+T^*-\phi^*\phi)^2}
\left|\frac{\partial W}{\partial \phi} \right|^2
\end{equation}
and no supersymmetry breaking masses arise in the scalar sector. Furthermore
the gravitino mass is not fixed, which can be arbitrarily heavy
or light at this level. Thus the no-scale model is named after this property.
Non-trivial dependence of gauge kinetic functions on the field $T$ yields
non-vanishing gaugino masses in this case.

The no-scale structure  appears when one considers a Calabi-Yau
compactification of weakly coupled $E_8 \times E_8$ heterotic string 
theory. If one focuses
on the overall modulus field whose scalar component represents the
overall size of the compactified space, then one finds 
\cite{Witten}
\begin{equation}
  K=-\ln(S+S^*)-3\ln(T+T^*-\phi^*\phi),
\label{eq:string}
\end{equation}
where $S$ is the dilaton field and $T$ is the overall modulus
field. The superpotential in this case generally depends on the these
fields $S$ and $T$. Now if $T$ dominates the SUSY breaking, then one
finds that the soft SUSY breaking scalar mass as well as a
trilinear scalar coupling ($A$ term) vanishes as the vacuum energy,
{\it i.e.} the vacuum expectation value of the scalar potential,
vanishes. Note that the $T$ dominant SUSY breaking occurs when the
gaugino condensation triggers the SUSY breaking.

The same structure was also obtained for the heterotic M-theory \cite{het-M}
which
corresponds to the strong coupling regime of the heterotic string
theory, but this time the fields $S$ and $T$ have physically different
meanings.  In both the weak-coupling and strong-coupling cases, one
has to keep in mind that quantum corrections may alter the form of the
K\"ahler potential (\ref{eq:string}).

Severe FCNC constraints on superparticle masses may suggest
that the hidden sector and the observable sector are in some way
separated from each other in the K\"ahler potential. An assumption
often taken along this line of reasoning is the separation of the two sectors in
the K\"ahler potential itself, namely the K\"ahler potential is a sum
of the contributions from the two sectors. This Ansatz will generate
the superparticle mass spectrum of the well-known minimal supergravity
model and  non-zero scalar masses arise. It may be, however, 
more natural to consider the same separation in the superspace density in the
supergravity Lagrangian \cite{IKYY},  before making Weyl
transformations to obtain the Einstein-Hilbert action for gravity part. 
This spirit indeed leads the form of the K\"ahler potential in 
Eq.~(\ref{eq:no-scale-structure}). In this case and in the string cases,
the gaugino masses become non-zero, provided that the hidden sector
couples to the gauge multiplets via the gauge kinetic functions.

Recently it has been pointed out that the 
form (\ref{eq:no-scale-structure}) is  naturally realized in a 
five-dimensional setting with two separated
3-branes \cite{Randall-Sundrum,Luty-Sundrum}. 
Consider the five-dimensional supergravity on $R^4 \times S^1/Z_2$.
The geometry has two four-dimensional boundaries, {\it i.e.} 3-branes.
Suppose that the hidden sector is on one of the 3-branes and the observable
sector is on the other. Now a dimensional reduction of the theory
yields, in four dimensions,  the following form of the K\"ahler potential
\begin{equation}
   K=-3\ln(T+T^* +f_{hid}(z,z^*) +f_{obs}(\phi,\phi^*)),
\end{equation}
where this time the real part of $T$ stands for the length of the 
compactified fifth dimension. 

In the brane separation scenario, the two sectors are really split
geometrically and thus not only the scalar masses, but also the
gaugino masses vanish. Therefore one needs to seek for another
mechanism to mediate the SUSY breaking occurred in the hidden
sector. One way is to invoke superconformal anomaly to obtain
loop-suppressed soft masses \cite{Randall-Sundrum,GLMR}. 
This anomaly mediation is very appealing,
albeit its minimal version has negative masses squared for
sleptons. Many attempts to build realistic models have been made \cite{AMSB}, 
and
the superparticle masses obtained are in general different from those from
the no-scale boundary conditions. In ref.~\cite{Nomura-Yanagida}, a new 
$U(1)$ gauge interaction is assumed to play a role of the mediator of 
the SUSY breaking. The resulting mass pattern is similar to that of 
gauge-mediated SUSY breaking.  On the other hand, if the SM gauge sector
lives in the bulk, then the gauginos can play a role of the SUSY-breaking
messenger \cite{gaugino-mediation} 
and the resulting mass spectrum of the superparticles exhibits the
no-scale structure with non-vanishing gaugino masses, which is given at
the scale of (the inverse of) the length of the fifth-dimension.

\section{Minimal Scenario}

In this section, we would like to discuss phenomenological consequences
of the minimal no-scale scenario which has been mainly studied in
the literature. The soft SUSY breaking masses in the minimal case are
parameterized by:
\begin{itemize}
\item  vanishing scalar masses:  $m_0=0$
\item  vanishing trilinear scalar couplings:  $A=0$
\item  non-zero Higgs mixing masses:  $B$
\item  non-zero universal gaugino masses:  $M_{1/2}$
\end{itemize}
Note that these values are given at the GUT scale $M_{GUT}\simeq 2\times
10^{16}$ GeV. In addition to these soft masses, we assume a non-zero
supersymmetric higgsino mass, $\mu$. These masses at the weak scale are
obtained by solving renormalization group equations. Given $M_{1/2}$,
requiring the correct electroweak symmetry breaking relates $B$ and $\mu$
with the $Z$ boson mass $m_Z$ and the ratio of the two Higgs vacuum
expectation values $\tan \beta$ as in the usual manner.


At first we roughly estimate the mass spectrum of superparticles
when the Yukawa effects and the left-right mixing effects are neglected.
The bino, wino and gluino masses at the weak scale are given 
by one parameter $M_{1/2}$ (in the following we set the renormalization point 
to be 500 GeV),
\begin{eqnarray}
  M_1^2 \simeq 0.18 M_{1/2}^2, \qquad  M_2^2 \simeq 0.69 M_{1/2}^2, \qquad
  M_3^2 \simeq 7.0 M_{1/2}^2 ~.
\end{eqnarray}
The soft SUSY breaking masses of scalars in the first-two generations are 
also determined by one parameter $M_{1/2}$.
\begin{eqnarray}
  \tilde{m}^2_{u_L} &\simeq& 5.8 M_{1/2}^2 + 0.35 m^2_Z \cos 2 \beta  \\
  \tilde{m}^2_{d_L} &\simeq& 5.8 M_{1/2}^2 - 0.42 m^2_Z \cos 2 \beta  \\
  \tilde{m}^2_{u_R} &\simeq& 5.4 M_{1/2}^2 + 0.15 m^2_Z \cos 2 \beta \\
  \tilde{m}^2_{d_R} &\simeq& 5.4 M_{1/2}^2 - 0.077 m^2_Z \cos 2 \beta 
\end{eqnarray}
\begin{eqnarray}
  \tilde{m}^2_{\ell_L} &\simeq&  0.51 M_{1/2}^2 -0.27 m^2_Z \cos 2 \beta   \\
  \tilde{m}^2_{\ell_R} &\simeq& 0.15 M_{1/2}^2 -0.23 m^2_Z \cos 2 \beta  \\
  \tilde{m}^2_{\nu} &\simeq& 0.51 M_{1/2}^2 +0.5 m^2_Z \cos 2 \beta
\end{eqnarray}
The terms proportional to $m_Z^2 \cos 2 \beta$ are 
$U(1)_Y$ D-term contributions.
From these equations, we find that bino and right-handed slepton are light.
When $M_{1/2} \gtrsim 2.8 m_Z \sim 260 ~\textrm{GeV}$, 
$U(1)_Y$ D-term contribution becomes small, and then 
the charged right-handed slepton becomes the LSP,
and this scenario contradicts to cosmological observations.

\begin{figure}[tbp]  
  \begin{center}
    \includegraphics[height=10cm,width=12cm,keepaspectratio,clip]
    {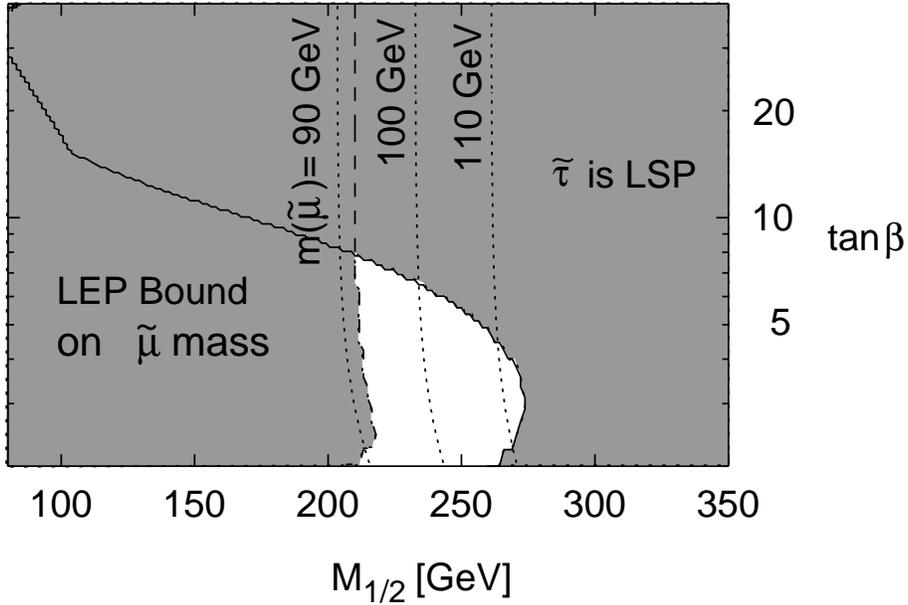}
 \end{center}
 \caption{Allowed region of the minimal no-scale scenario. The horizontal 
   axis is the universal gaugino mass at the GUT scale $M_{1/2}$ and the 
   vertical axis is $\tan \beta$.  In the region above the solid line
   $\tilde{\tau}$ is the LSP and it should be cosmologically excluded. 
   The left side of the dashed line is excluded by smuon searches 
   by the LEP experiments at $\sqrt{s}=202$ GeV. 
   We also show the contour of right-handed smuon mass. }
 \label{fig:minimalbound}
\end{figure}

In Fig.~\ref{fig:minimalbound}, we show the numerical result.
The region above the solid line is excluded cosmologically 
since charged stau is the LSP.
For $\tan \beta \lesssim 10$ where left-right mixing effect is negligible,
the region $M_{1/2} \gtrsim 260 ~\textrm{GeV}$ is excluded 
as we estimate above.
For $\tan \beta \gtrsim 10$, 
since left-right mixing effect makes stau mass lighter ,
the constraint becomes stronger. 
In Fig.~\ref{fig:minimalbound}, 
we also show the value of the right-handed smuon mass.
From the cosmological constraint, we find that the right-handed smuon 
must be lighter than about 120 GeV.
 
On the other hand, the LEP experiments at $\sqrt{s} = 202 ~
\textrm{GeV}$ provide rather strong lower bound on slepton masses
\cite{SUSY2000LEP}.  For smuon, except near the threshold, the cross
section for smuon pair production $\sigma ( e^+ e^- \to
\tilde{\mu}^+_R \tilde{\mu}^-_R )$ must be smaller than 0.05 pb to
survive the smuon searches at LEP.  Here we impose that $\sigma ( e^+
e^- \to \tilde{\mu}^+_R \tilde{\mu}^-_R ) \leq 0.05 ~\textrm{pb}$ for
$m_{\tilde{\mu}_R} \leq 98 ~\textrm{GeV}$ and $m_{\tilde{\chi^0_1}}
\leq 0.98 m_{\tilde{\mu}_R} - 4.1 ~\textrm{GeV}$.  This constraint
excludes the left side of the dashed line in the
Fig.~\ref{fig:minimalbound}.  Combining these two constraints, we
conclude that the no-scale scenario with the universal gaugino masses
is allowed only for $\tan \beta \lesssim 8$ and $ 210 ~
\textrm{GeV}\lesssim M_{1/2} \lesssim 270 ~\textrm{GeV}$.

\section{Case of Non-Universal Gaugino Masses}
In this section, we consider modifications of the minimal boundary conditions
discussed in the previous section, and argue that slight modifications will
drastically change phenomenological consequences.

The reason of the very constrained superparticle mass spectrum in the
minimal case is the degeneracy of the bino mass and those of the
right-handed sleptons. The degeneracy is resolved if one considers the
renormalization group effects above the GUT scale \cite{KMY,PolonskyPomarol,Schmaltz-Skiba}. 
The point is that the right-handed slepton multiplets
belong to 10-plets in the minimal choice of the matter representations
in the $SU(5)$ GUT, and the large group factor in the gauge loop
contributions yields large positive corrections to the slepton masses.
We should note, however, that in some realistic models to attempt to
explain the masses of quarks, leptons and neutrinos, matter multiplets
in different generations are often taken to be in different 
representations of the GUT groups \cite{SatoYanagida},
and then the renormalization group effects would violate the mass
degeneracy among the different generations, which might cause unacceptably
large FCNCs. 

Secondly the stau can be the lightest superparticle in the SSM sector if
it is not stable. This is indeed the case when R-parity is violated or
there exists another superparticle such as a gravitino out of the SSM sector 
which is lighter than the stau \cite{light-gravitino}. 

Another possibility is to relax the universality of the gaugino masses.
In the rest of this section, we will discuss this case in detail. 
In the next subsection, we shall review various possibilities to
realize non-universal gaugino masses. In particular we will emphasize
that the non-universality of the gaugino masses does not conflict with the 
universality of the gauge couplings.  Then we will look into phenomenological
implications of the non-universality. 

\subsection{Examples of Non-Universal Gaugino Masses}

Once the gaugino masses are given universal at some high energy scale
where the gauge groups are unified, it is shown that the gaugino mass relation
$M_1:M_2:M_3 \simeq 1:2:6$ holds at low energy, irrespective of the
breaking patterns of the GUT group \cite{KMY1,KMY}.
Here we review some mechanisms in which the gaugino masses are non-universal
from the beginning.  

In string models with simple Calabi-Yau compactification, the gauge
kinetic functions for the Standard Model gauge multiplets can be
written as \cite{BIM}
\begin{equation}
    f_i=S +\epsilon_i T
\end{equation}
where $i=1$, 2, 3 represent the 
three Standard Model gauge groups and $\epsilon_i$
are some coefficients of one-loop order 
determined by the details of the compactification.
If $\epsilon_i$ depends on a gauge group and the modulus field $T$ is 
dominantly responsible for the SUSY breaking, we will have the 
non-universal gaugino  masses:
\begin{equation}
   M_1 : M_2 : M_3 = \epsilon_1 : \epsilon_2 : \epsilon_3.
\end{equation} 
Here we would like to emphasize that large threshold corrections are
necessary for the string unification scenario in the weak
coupling regime where the string scale is more than one order of magnitude
larger than the naive GUT scale, and thus appearance of the non-universal
$\epsilon_i$ terms seems to be requisite. Note again that the 
K\"ahler potential may receive quantum 
corrections at the same order and the no-scale structure may be distorted. 

The non-universality of the gaugino masses can be achieved in the conventional
GUT approaches. Suppose that the gauge kinetic functions are written in
the following form \cite{GUT-nonuniversal}:
\begin{equation}
          f=c +\Sigma Z   \label{eq:gauge-kinetic-function-GUT}
\end{equation}
where $c$ is a universal constant, $ \Sigma$ is a field which breaks the
GUT group to the SM group, and $Z$ is assumed to break the SUSY. The first
term respects the GUT symmetry and thus universal for all SM gauge groups,
while the second term is a symmetry breaking part which depends on each
SM group. As for the gauge couplings, the first term gives a dominant
contribution and hence the gauge couplings are unified up to small
non-universal effects from the second term. On the other hand, the
gaugino masses are assumed to come from the second term in 
Eq.~(\ref{eq:gauge-kinetic-function-GUT}). 
They are proportional
to the vacuum expectation value of $\Sigma$ and thus  non-universal.
The form of Eq.~(\ref{eq:gauge-kinetic-function-GUT}) can also be obtained through
GUT threshold corrections to the gauge kinetic functions 
\cite{Hisano-Goto-Murayama}.

Non-universal  gaugino masses can also be realized in
scenarios of product GUTs \cite{HIY} where the gauge group has the structure of
$G_{GUT} \times G_H$ and the Standard Model gauge groups are obtained
as diagonal subgroups of the two product groups. The idea of the
product GUTs provides an elegant solution to the triplet-doublet
splitting problem in the Higgs sector based on the missing doublet
mechanism. The gauge coupling unification achieves if the gauge
couplings of the $G_H$ group are sufficiently large, 
while contributions to the gaugino masses from the $G_H$ sector are generally
sizable and destroy their universality. 
\cite{non-universal-product-GUT}.

The flipped $SU(5)$ is another example where the non-universality of the
gaugino masses naturally arises \cite{MizutaNgYamaguchi}. The gauge 
group is $SU(5) \times U(1)$ and thus even if the $SU(5)$ part gives
a universal contribution the gaugino mass from  $U(1)$ in general gives 
a different mass, violating the universality of the $U(1)_Y$ gaugino mass
with the rest two.  

In summary, the non-universality of the gaugino masses is not a
peculiar phenomenon even in the light of the gauge coupling unification.
Motivated by this observation, we will discuss its 
phenomenological consequences. 

\subsection{Phenomenological Implications}
In this subsection we discuss some phenomenological implications of 
non-universal gaugino masses.
At the cutoff scale, all scalar masses are vanishing as in the minimal case, 
while the bino, winos and gluinos possess nonzero masses $M_{1,0}$, $M_{2,0}$, 
and $M_{3,0}$ , respectively,
and now they are no longer degenerate in general.
The soft SUSY breaking mass parameters at the weak scale are obtained 
by solving the RGEs. 
In this paper we use the one-loop level RGEs.
With the soft SUSY breaking masses, we evaluate the physical masses 
using the tree-level potential. 
We also obtain the value of $\mu$ from the electroweak symmetry breaking 
condition with tree-level Higgs potential.

Before showing numerical results,
we discuss the mass spectrum of superparticles when the Yukawa effects to 
the RG evolutions and left-right mixings are neglected. 
Relations of the gaugino masses at the GUT scale $M_{GUT}$ and the 
electroweak scale $M_{EW}$ are
\begin{equation}
  M_1^2 \simeq 0.18 M_{1,0}^2   ~, \quad
  M_2^2 \simeq 0.69 M_{2,0}^2   ~, \quad
  M_3^2 \simeq 7.0 M_{3,0}^2   ~.
  \label{eq:gauginomass@EW}  
\end{equation}
Neglecting effects of Yukawa interaction, 
the masses squared of sfermions at the weak scale are evaluated to be
\begin{eqnarray}
  \tilde{m}^2_{u_L} &\simeq& 5.4 M_{3,0}^2 +0.47 M_{2,0}^2 
    + 4.2 \times 10^{-3} M_{1,0}^2  + 0.35 m^2_Z \cos 2 \beta  \\
  \tilde{m}^2_{d_L} &\simeq& 5.4 M_{3,0}^2 +0.47 M_{2,0}^2 
    + 4.2 \times 10^{-3} M_{1,0}^2  - 0.42 m^2_Z \cos 2 \beta  \\
  \tilde{m}^2_{u_R} &\simeq& 5.4 M_{3,0}^2 + 0.066 M_{1,0}^2
    + 0.15 m^2_Z \cos 2 \beta \\
  \tilde{m}^2_{d_R} &\simeq& 5.4 M_{3,0}^2 + 0.017 M_{1,0}^2
    - 0.077 m^2_Z \cos 2 \beta 
\end{eqnarray}
\begin{eqnarray}
  \tilde{m}^2_{\ell_L} &\simeq&  0.47 M_{2,0}^2 + 0.037 M_{1,0}^2
  -0.27 m^2_Z \cos 2 \beta
  \label{eq:ellL} \\
  \tilde{m}^2_{\ell_R} &\simeq& 0.15 M_{1,0}^2 -0.23 m^2_Z \cos 2 \beta
  \label{eq:ellR} \\
  \tilde{m}^2_{\nu} &\simeq& 0.47 M_{2,0}^2 + 0.037 M_{1,0}^2
  +0.5 m^2_Z \cos 2 \beta         ~.
  \label{eq:nu}
\end{eqnarray}
From the above equations, we find that if $M_{1,0} \gtrsim 2.0 M_{2,0}$,
$\tilde{m}^2_{\ell_R}$ is heavier than $M_1^2$, $M_2^2$ , 
$\tilde{m}^2_{\ell_L}$ and $\tilde{m}^2_{\nu}$.
Notice that the mass of the charged left-handed slepton is heavier than 
the mass of the neutral sneutrino 
because $\cos 2 \beta \leq 0$ for $\tan \beta \geq 1$ .
On the other hand, for $M_{1,0}/M_{2,0} \gtrsim 2.5$, the wino mass tends to be 
lighter than the sneutrino mass.
Hence we expect that sneutrino can be LSP when 
$2 \lesssim M_{1,0} /M_{2,0} \lesssim 2.5$, 
and wino-like neutralino can be LSP when $M_{1,0} /M_{2,0} \gtrsim 2.5$.

Next, we consider how $\mu$ affects the mass spectrum of the superparticles.
The value $\mu$ is determined by minimizing the Higgs potential.
At the tree-level, $\mu$ is calculated in terms of the soft SUSY breaking 
masses of the Higgses and $\tan \beta$,
\begin{equation}
  \mu^2 = \frac{ \tilde{m}^2_{H_d} - \tilde{m}^2_{H_d} \tan^2 \beta}
   {\tan^2 \beta - 1} - \frac{1}{2} m^2_Z   ~.
\end{equation}
In order to obtain the value of $\tilde{m}^2_{H_d}$ and $\tilde{m}^2_{H_u}$,
we have to include the Yukawa interaction.
For the moment we consider the low $\tan \beta$ region, i.e.,
we take only the top Yukawa coupling into account and neglect the bottom and 
tau Yukawa couplings for simplicity.
In this case $\tilde{m}^2_{H_d} = \tilde{m}^2_{\ell_L}$ and 
we can obtain an analytic solution for the RGE of $\tilde{m}^2_{H_u}$. 
For $\tan \beta = 10$ , $\mu$ is approximately 
\begin{eqnarray}
  \mu^2 &=& 2.1 M_{3,0}^2 - 0.22 M_{2,0}^2 - 0.0064 M_{1,0}^2 
     + 0.0063 M_{1,0} M_{2,0} \nonumber \\ 
     & &+ 0.19 M_{2,0} M_{3,0} + 0.029 M_{3,0} M_{1,0}
    - \frac{1}{2} m_Z^2    ~.
\label{eq:tanb10mu}
\end{eqnarray}
From this equation, we find that the size of $\mu$ is strongly correlated with 
the size of the gluino mass $M_{3,0}$ and $|\mu|$ becomes large 
as $M_{3,0}$ increases.
Hence when $M_{3,0}$ is large enough, 
the left-right mixing in the slepton masses is important,
which makes one of the staus, $\tilde{\tau}_1$, lighter than sneutrino.
On the other hand if $M_{3,0}$ is small enough, $|\mu|$ becomes smaller than 
the mass of bino, wino, slepton and sneutrinos,
and then a higgsino-like neutralino can be the LSP.
Actually, for $\tan \beta = 10$, from eq. (\ref{eq:tanb10mu}) we find that 
$|\mu|$ is smaller than $M_2$ if $M_{3,0}/M_{2,0} \lesssim 0.5$ is satisfied.

In the non-universal case, not only the mass spectrum 
but also the mixing properties of the neutralinos are very different 
from those in the minimal case.
To see this we classify the lightest neutralino $\chi^0_1$ 
into five cases as follows.
$\tilde{\chi}^0_1$ is a linear combination of bino, wino and higgsinos and 
is written as 
\begin{equation}
      \tilde{\chi}^0_1 = (O_N)_{1B} \tilde{B} + (O_N)_{1W} \tilde{W} 
    + (O_N)_{1H_d} \tilde{H_d} + (O_N)_{1H_u} \tilde{H_u} ~,
\end{equation}
where $O_N$ is orthogonal matrix diagonalizing the neutralino mass matrix.
When $ | (O_N)_{1B} |^2 > 0.8$, $ | (O_N)_{1W} |^2 > 0.8$ or
$ |(O_N)_{1H_d}|^2 + |(O_N)_{1H_u}|^2 > 0.8$, we call these parameter region 
'bino region', 'wino region' or 'higgsino region', respectively.
When $ | (O_N)_{1B} |^2 < 0.8$ and $ | (O_N)_{1W} |^2 < 0.8$ and 
$ | (O_N)_{1B} |^2 + | (O_N)_{1W} |^2 > 0.8$ , 
we call the region 'bino-wino mixed region'.
The other parameter region is called 'mixed region'.

\begin{figure}[tbp]  
  \begin{center}
    \includegraphics[height=10cm,width=12cm,keepaspectratio,clip]
    {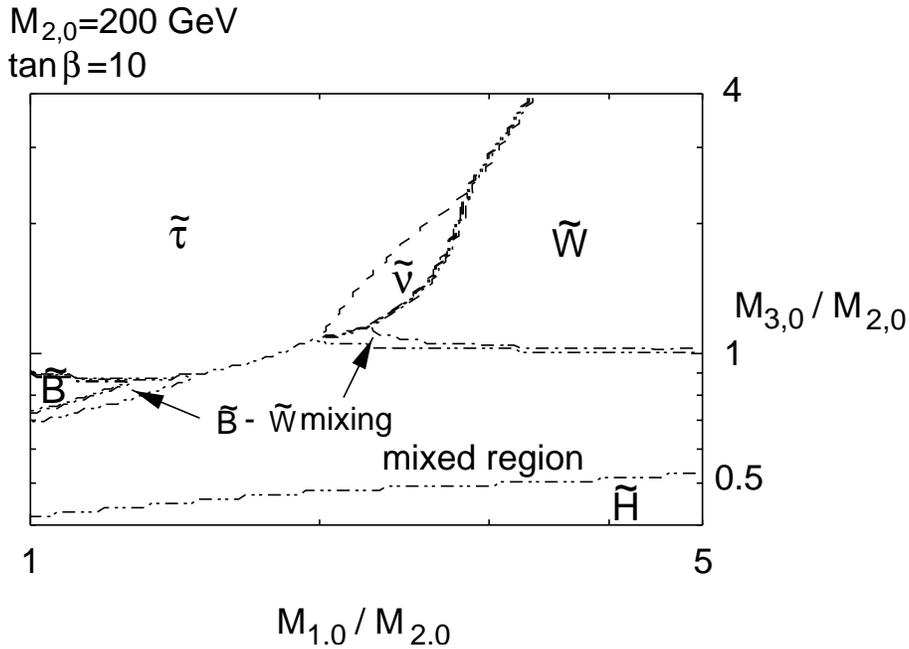}
  \end{center}
  \caption{The composition of the LSP in the $M_{1,0}/M_{2,0}-M_{3,0}/M_{2,0}$
    plane, for $M_{2,0}=200 \textrm{GeV}$ and $\tan \beta = 10$. 
    The classification of the neutralino LSP is given 
    in the text.}
\label{fig:LSPM2200tanb10MBCMGUT}
\end{figure}

In Fig. \ref{fig:LSPM2200tanb10MBCMGUT} we show the composition of the LSP 
when we relax the gaugino mass universality. 
Here we take $M_{2,0} = 200 \textrm{GeV}$, 
$\tan \beta = 10$ and $\textrm{sgn}(\mu) = +1$.
Recall that for the universal gaugino masses at the GUT scale, 
the LSP is the lighter stau and this parameter set is excluded.
Once we relax the universality, however, 
we see that the situation drastically changes,
and the composition of the LSP behaves as we have discussed with the
 approximate 
expressions eq.(\ref{eq:gauginomass@EW}) - eq.(\ref{eq:nu}).
The lightest neutralino can be the LSP in a large parameter region , 
and furthermore unlike the universal case, 
it can be wino-like, higgsino-like or admixture of them as well as bino-like. 
When $M_{1,0}/M_{2,0} \gtrsim 2.5$ and $M_{3,0}/M_{2,0} \gtrsim 1$ 
the wino is the LSP.
And as the ratio $M_{3,0}/M_{2,0} $ decreases, $|\mu|$ 
becomes comparable to $M_1$ and $M_2$ and the lightest neutralino is the 
admixture of bino, wino and higgsinos.
Further $M_{3,0}/M_{2,0} $ becomes smaller than about 0.5, 
the dominant component of the lightest neutralino is higgsino.
Also we find in the region $2 \lesssim M_{1,0}/M_{2,0} \lesssim 2.5$, 
the tau sneutrino is indeed the LSP. 
And we find that when $M_{3,0}/M_{2,0}$ is larger than 2, i.e., 
$|\mu|$ is large and so is the left-handed and right-handed stau mixing, 
sneutrino can not be the LSP, and stau is LSP even when $M_{1,0}/M_{2,0}$ is 
bigger than 2.5 -- 3.

\begin{figure}[tbp]  
  \begin{center}
    \includegraphics[height=10cm,width=12cm,keepaspectratio,clip]
    {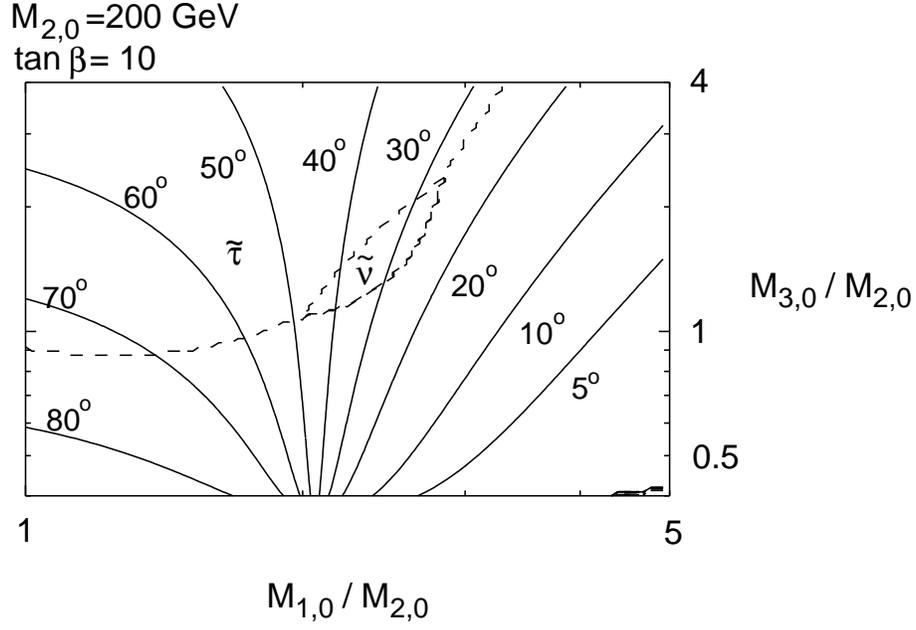}
  \end{center}
  \caption{Mixing angle $\theta_{\tau}$ in the $M_{1,0}/M_{2,0} - 
    M_{3,0}/M_{2,0}$ plane for $M_{2,0}=200 \textrm{GeV}$ and 
    $\tan \beta = 10$. We also show the region where stau and tau 
    sneutrino are the LSP. The definition of the mixing angle is given 
    in the text.}
\label{fig:thetatau}
\end{figure}

In the non-universal case sfermions, as well as neutralinos and charginos, 
show variety of mass spectrum.
From eq. (\ref{eq:ellL}) and eq. (\ref{eq:ellR}), we find that when 
$M_{1,0} /  M_{2,0} \gtrsim 2$ left-handed sfermions are smaller than 
right-handed sfermions in contrast to the universal case.
For stau, the mixing angle of $\tilde{\tau}_L$ and $\tilde{\tau}_R$ 
also depends on this ratio.
In Fig. \ref{fig:thetatau} we show the behavior of this mixing angle 
$\theta_{\tau}$ in the 
$M_{1,0} /  M_{2,0} - M_{3,0} /  M_{2,0}$ plane,
where $\theta_{\tau}$ is defined such that the lighter stau $\tilde{\tau}_1$
is written as $\tilde{\tau}_1 = \cos \theta_{\tau} \tilde{\tau}_L + \sin \theta_{\tau} \tilde{\tau}_R$.
Around $M_{1,0}/M_{2,0} \simeq 2$, the mass of the right-handed stau is as heavy as that of the left-handed stau, and they mix maximally 
($\theta_\tau = 40 - 50$) as expected. 
Also masses of squarks strongly depend on $M_3$ , and thus 
mass relations between squarks and sleptons drastically change.
As we shall see later, some of the squarks can be lighter than the sleptons. 

\begin{figure}[tbp]  
  \begin{center}
    \includegraphics[height=10cm,width=12cm,keepaspectratio,clip]
    {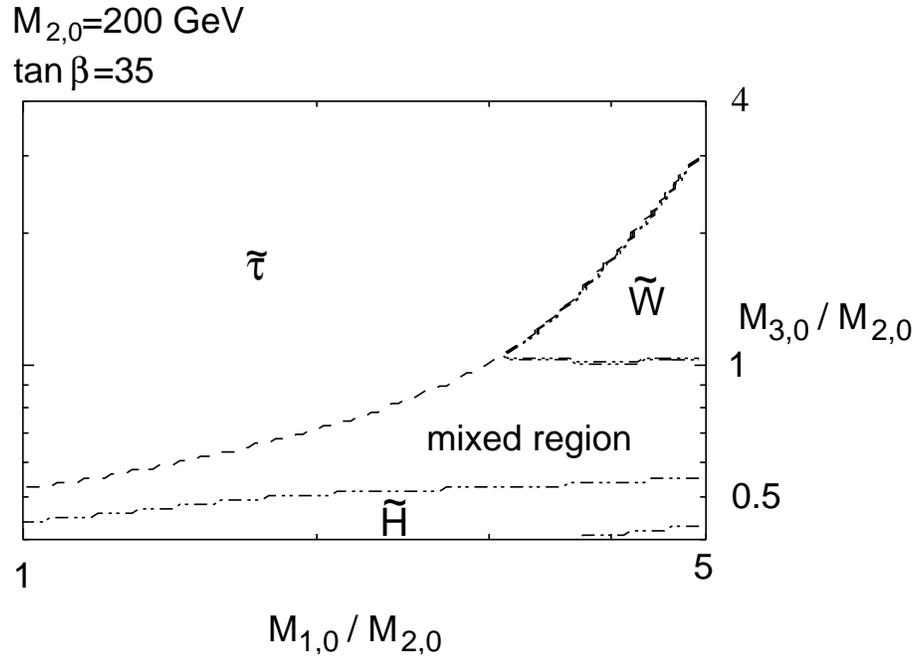}
  \end{center}
  \caption{The same as Fig.~\ref{fig:LSPM2200tanb10MBCMGUT}, 
    but  for $M_{2,0}=200 \textrm{GeV}$ and $\tan \beta = 35$. }
\label{fig:LSPM2200tanb35MBCMGUT}
\end{figure}

\begin{figure}[tbp]  
  \begin{center}
    \includegraphics[height=10cm,width=12cm,keepaspectratio,clip]
    {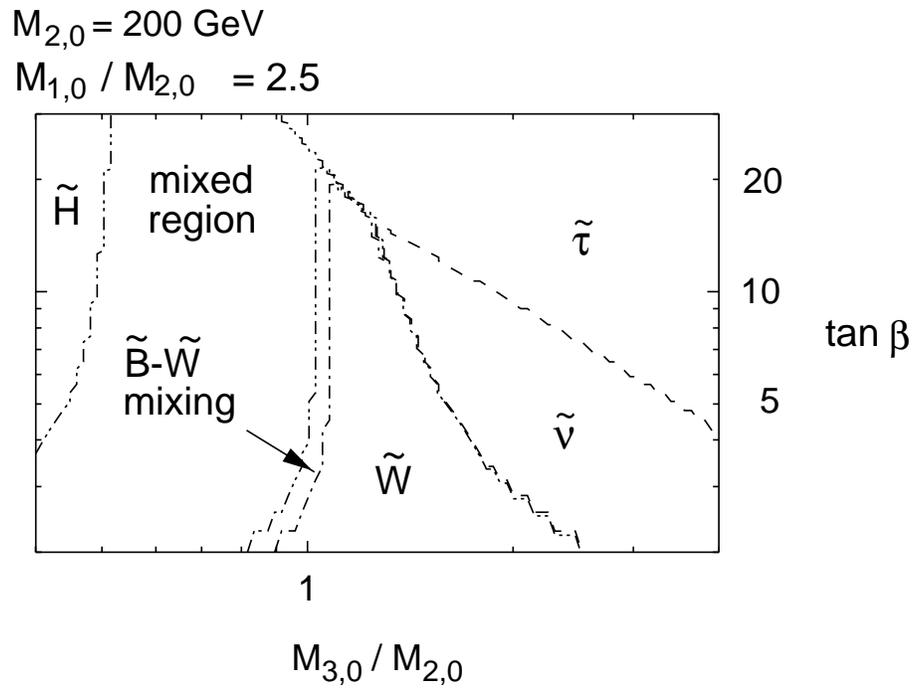}
  \end{center}
  \caption{ The composition of the LSP in the $M_{3,0}/M_{2,0} - \tan \beta$ 
    plane, for $M_{2,0} = 200 \textrm{GeV}$ and $M_{1,0}/M_{2,0} = 2.5$ .}
  \label{fig:LSPM2200MBCMGUTRat122.5}
\end{figure}

In Fig.~\ref{fig:LSPM2200tanb35MBCMGUT} we show the same graph 
as Fig.~\ref{fig:LSPM2200tanb10MBCMGUT} except for $\tan \beta = 35$. 
In this case the Yukawa interaction and the left-right mixing make 
the stau mass lighter.
In fact, although wino-like, higgsino-like and mixed neutralino is LSP 
in large parameter region, 
the sneutrino can not be the LSP.
To see the relation among the stau mass, the tau sneutrino mass and $\tan \beta$,
we plot in Fig.~\ref{fig:LSPM2200MBCMGUTRat122.5} the composition of the LSP 
in the $M_{3,0}/M_{2,0} - \tan \beta$ plane,
fixing $M_{3,0}/M_{2,0} = 2.5$.
This figure shows that the tau sneutrino can be the LSP 
when $\tan \beta \lesssim 15$ where the left-right mixing is not so sizable.
We have checked that these features are insensitive to the signs of $\mu$ and 
gaugino masses.

We shall next investigate the mass spectrum of superparticles in detail,
by choosing some representative parameter sets, 
and discuss phenomenology for each parameter set.
The points we choose are listed in Table \ref{tab:gauginomassGUT}. 
In table \ref{tab:N1component} we show contamination of $\tilde{\chi}^0_1$
for each point.
At the points A and E, the LSP is the wino-like neutralino.
At the points B, C and E the LSP is the higgsino-like neutralino.
And at the point D the tau sneutrino is the LSP.
In table \ref{tab:spectrum} we list the mass spectrum of superparticles.

The wino-like neutralino is the LSP when $M_{1,0}/M_{2,0} \gtrsim 2$ and 
$M_{3,0}/M_{2,0} \gtrsim 1$ .
In the wino-like neutralino LSP case, the lighter chargino and 
the lightest neutralino are highly degenerate generally.
This character and the resulting phenomenology has been studied in 
\cite{MizutaNgYamaguchi,1Loopmass,ChenDress96,FengMoroietal99,FengMoroi00,GunionMrenna00}. 
On top of this our scenario also predicts that the right-handed sfermions are 
heavier than the left-handed ones because of the inequality 
$M_{1,0}/M_{2,0} \gtrsim 2$ ,
and colored superparticles are heavier than other superparticles because of the 
inequality $M_{3,0}/M_{2,0} \gtrsim 1$
(see the list for the points A and E in the table \ref{tab:spectrum}) .
The former may be an interesting feature. 
Anomaly-mediated SUSY breaking (AMSB) scenario also predicts the wino-like LSP.
However in the minimal AMSB model where universal mass is added to all scalars
to avoid negative slepton masses squared, 
the left-handed and right-handed sleptons in the first two generations tend 
to be degenerate \cite{FengMoroi00}.
Thus we can distinguish two scenarios with the wino LSP, 
the no-scale scenario with non-universal gaugino masses and the minimal AMSB,
by measuring these slepton masses.

The higgsino-like neutralino is the LSP when $M_{3,0}/M_{2,0} \lesssim 0.5$
regardless of $M_{1,0}/M_{2,0}$ .
In the higgsino-like neutralino LSP case, the mass deference between 
higgsino-like neutralino LSP and chargino NLSP is generally small.
The resulting phenomenology have been studied in 
\cite{MizutaYamaguchi,GiudicePomarol96,DressNojiri97}.
Furthermore in our case, the sleptons are as heavy as the squarks due to the 
inequality $M_{3,0}/M_{2,0} \lesssim 0.5$.
Especially the lighter stop and sbottom can be lighter than 
some of the sleptons.
Actually, at the points B, C, and F, the lighter stop is comparable 
to the slepton masses, and all superparticle masses are below 400 - 450 GeV.

In the non-universal scenario, the tau sneutrino can also be the LSP 
when $2 \lesssim M_{1,0}/M_{2,0} \lesssim 2.5$,
$1 \lesssim M_{3,0}/M_{2,0} \lesssim 5$ and $\tan \beta \lesssim 15$.
From the first inequality, we find that the mass difference between 
left-handed and right-handed squarks in the first two generations is small,
and the left-right mixing angle of the stau is big 
as shown in Fig. \ref{fig:thetatau}.

\section{Conclusions}

In this paper, we have revisited the no-scale scenario where the
vanishing SUSY-breaking scalar masses and trilinear scalar couplings
are given at the GUT scale. When the gaugino masses are given universal,
the renormalization group analysis implies that the bino mass and the 
right-handed slepton masses are close to each other.  This degeneracy
leads an upperbound of the LSP mass around 120 GeV: above it the 
LSP would be the charged stau, which must be excluded cosmologically.
Furthermore, the negative results of the slepton searches at LEP200 
already excluded a large portion of the parameter space including a
large $\tan \beta$ region, leaving $\tan \beta \lesssim 8$.

We next considered various ways out to avoid the aforementioned severe
constraints. Among them, we concentrated on the case of the
non-universal gaugino masses. In fact the non-universality of the
gaugino masses is by no means a peculiar phenomenon, rather it is
realized in various scenarios, including some approaches to grand
unification. We investigated some phenomenological implications of the
no-scale model with the non-universal gaugino masses. We found that
there is no longer severe constraint on the superparticle masses and
the mass spectrum of the superparticle has much richer structure. In 
particular, the LSP can be the wino-like neutralino, the higgsino-like
neutralino, or even the sneutrino.  We also found that unlike the conventional
universal gaugino mass case, the left-handed slepton masses can be lighter
than the right-handed slepton masses.  We expect that resulting collider
signatures with these features will be quite different from the usual 
scenario with the universal gaugino masses. Further studies along
this direction should be encouraged.

\section*{acknowledgment}
We would like to thank Y. Nomura, T. Moroi and Y. Yamada for useful
discussions.  This work was supported in part by the Grant-in-Aid for
Scientific Research from the Ministry of Education, Science, Sports,
and Culture of Japan, on Priority Area 707 "Supersymmetry and Unified
Theory of Elementary Particles", and by the Grants-in-Aid No.11640246
and No.12047201.

\begin{table}
  \begin{center}
    \begin{tabular}{|c|cccccc|}   
               &Point A&Point B&Point C&Point D&Point E&Point F \\  \hline
  $M_{1,0}$    &  800  & 1000  &  400  &  500  &  800  &  600   \\
  $M_{2,0}$    &  200  &  250  &  200  &  200  &  200  &  200   \\
  $M_{3,0}$    &  400  &  125  &  100  &  300  &  300  &  100   \\
  $\tan \beta$ &   10  &   10  &   10  &   10  &   35  &   35   \\  
    \end{tabular}
  \end{center}
  \caption{Gaugino masses at the GUT scale for each point.
    All dimensionful parameters are given in the GeV unit.}
\label{tab:gauginomassGUT}
\end{table}
\begin{table}
  \begin{center}
    \begin{tabular}{|c|cccccc|} 
                &Point A&Point B&Point C&Point D&Point E&Point F \\  \hline
$(O_{N})_{1B}$  &-0.017 &0.0835 & 0.241 & 0.092 &-0.022 & 0.126  \\
$(O_{N})_{1W}$  & 0.987 &-0.478 &-0.457 &-0.967 & 0.973 &-0.445  \\
$(O_{N})_{1H_d}$&-0.149 &0.689  & 0.710 & 0.219 &-0.213 & 0.729  \\
$(O_{N})_{1H_u}$& 0.054 &-0.539 &-0.479 &-0.096 & 0.084 &-0.504  \\
   \end{tabular}
 \end{center}
 \caption{Components of the lightest neutralino 
   $\tilde{\chi}^0_1$ which 
is a linear combination of bino, wino and higgsinos, 
   $ \tilde{\chi}^0_1 = (O_{N})_{1B} \tilde{B} + 
   (O_{N})_{1W} \tilde{W} + (O_{N})_{1H_d} \tilde{H_d} + 
   (O_{N})_{1H_u} \tilde{H_u}$. }
\label{tab:N1component}
\end{table}
\begin{table}
  \begin{center}
    \begin{tabular}{|c|cccccc|} 
     particle       &Point A&Point B&Point C&Point D&Point E&Point F \\  \hline
     $\chi^0_1$     & 160   & 106   & 70    & 156   & 159  &  72 \\
     $\chi^0_2$     & 336   & 152   & 126   & 209   & 332  &  120 \\
     $\chi^0_3$     & 594   & 248   & 169   & 444   & 438  &  202 \\
     $\chi^0_4$     & 603   & 430   & 222   & 457   & 453  &  267 \\
     $\chi^+_1$     & 160   & 113   & 81    & 157   & 159  &  81 \\ 
     $\chi^+_2$     & 602   & 253   & 216   & 457   & 449  &  212 \\  \hline
 
     $\tilde{u}_L$  & 929   & 336   & 263   & 701   & 702  &  265  \\
     $\tilde{d}_L$  & 932   & 345   & 275   & 706   & 707  &  277  \\
     $\tilde{u}_R$  & 941   & 384   & 249   & 700   & 718  &  274  \\
     $\tilde{d}_R$  & 925   & 316   & 237   & 692   & 697  &  244  \\
     $\tilde{\nu}$  & 196   & 250   & 144   & 155   & 196  &  167  \\
     $\tilde{e}_L$  & 212   & 262   & 164   & 174   & 212  &  185  \\
     $\tilde{e}_R$  & 312   & 389   & 161   & 198   & 312  &  236  \\  \hline

     $\tilde{t}_1$  & 742   & 233   & 164   & 538   &  544 &  176 \\
     $\tilde{t}_2$  & 925   & 409   & 339   & 721   &  709 &  338 \\
     $\tilde{b}_1$  & 855   & 293   & 230   & 646   &  602 &  200 \\
     $\tilde{b}_2$  & 922   & 317   & 252   & 691   &  676 &  252 \\
$\tilde{\nu_\tau}$  & 195   & 249   & 143   & 154   &  183 &  156 \\
 $\tilde{\tau}_1$   & 205   & 261   & 154   & 159   &  166 &  169 \\ 
 $\tilde{\tau}_2$   & 314   & 387   & 169   & 209   &  316 &  225 \\  \hline

     $\tilde{g}$    & 1053  & 329   & 263   & 790   &  790 &  263 \\ 
   \end{tabular}
 \end{center}
 \caption{Mass spectrum for each point.
   All values are given in the GeV unit.}
\label{tab:spectrum}
\end{table}

\begin{references}

\bibitem{no-scale-model}
J. Ellis, C. Kounnas and D.V. Nanopoulos, Nucl. Phys. B247, 373 (1984).

\bibitem{Witten}
E. Witten, Phys. Lett. B155, 151 (1985).


\bibitem{IKYY}
K. Inoue, M. Kawasaki, M. Yamaguchi and T. Yanagida, 
Phys. Rev. D45, 328 (1992).

\bibitem{Randall-Sundrum}
L. Randall and R. Sundrum, Nucl. Phys. B557, 79 (1999). 

\bibitem{Luty-Sundrum}
M. Luty and R. Sundrum, hep-th/9910202. 

\bibitem{KLNPY}
S. Kelly, J.L. Lopez, D.V. Nanopoulos, H. Pois and K.-J. Yuan,
Phys. Lett. B273, 423 (1991).

\bibitem{Drees-Nojiri}
M. Drees and M.M. Nojiri, Phys. Rev. D45, 2482 (1992).

\bibitem{het-M}
T. Banks and M. Dine, Nucl. Phys. B505, 445 (1997);
H.P. Nilles, M. Olechowski and M. Yamaguchi, Phys. Lett. B415 (1997) 24;
Nucl. Phys. B530, 43 (1998);
Z. Lalak and S. Thomas, Nucl. Phys. B515, 55  (1998); 
A. Lukas, B.A. Ovrut and D. Waldram, Nucl. Phys. B532, 43 (1998);  
Phys. Rev. D57, 7529 (1998); 
K. Choi, H.B. Kim and C. Mu\~noz, Phys. Rev. D57, 7521 (1998). 

\bibitem{GLMR}
G.F. Giudice, M.A. Luty, H. Murayama and R. Rattazzi, JHEP 9812, 027 (1998). 

\bibitem{AMSB}
See for example, A. Pomarol and R. Rattazzi, JHEP 9905, 013 (1999);
Z. Chacko, M.A. Luty, I. Maksymyk and E. Pont\'on,  JHEP 0004, 001 (2000);
E. Katz, Y. Shadmi and Y. Shirman, JHEP 9908, 015 (1999);
K.-I. Izawa, Y. Nomura and T. Yanagida, Prog. Theor. Phys. 102, 1181 (1999);
I. Jack and D.R.T. Jones,  Phys. Lett. B482, 167 (2000).

\bibitem{Nomura-Yanagida}
Y. Nomura and T. Yanagida, hep-ph/0005211.

\bibitem{gaugino-mediation}
D.E. Kaplan, G.D. Kribs and M. Schmaltz,  hep-ph/9911293;
Z. Chacko, M. Luty, A.E. Nelson and E. Pont\'on, JEHP 0001, 003 (2000);
M. Schmaltz and W. Skiba, hep-ph/0001172.

\bibitem{SUSY2000LEP}
G. Ganis, 'Standard SUSY at LEP', talk presented in SUSY2K, 8th Int. Conf. on 
Supersymmetry in Physics, CERN, Geneva, Switzerland, June 26 - July 1, 2000

\bibitem{KMY}
Y. Kawamura, H. Murayama and M. Yamaguchi, Phys. Rev. D51, 1337 (1995).

\bibitem{PolonskyPomarol} 
N. Polonsky and A. Pomarol, Phys. Rev. D51, 6532 (1995)

\bibitem{Schmaltz-Skiba}
M. Schmaltz and W. Skiba, hep-ph/0004210.

\bibitem{SatoYanagida}
See for example, J. Sato and T. Yanagida,
Phys. Lett. B430, 127 (1998);
Y. Nomura and T. Yanagida, Phys. Rev. D59, 017303 (1999);
M. Bando and T. Kugo, Prog. Theor. Phys. 101, 1313 (1999).

\bibitem{light-gravitino}
T. Moroi, H. Murayama and M. Yamaguchi, Phys. Lett. B303 (1993)
289;
T. Gherghetta, G.F. Giudice and A. Riotto, Phys. Lett. B446 (1999) 28;
T. Asaka, K. Hamaguchi and K. Suzuki, hep-ph/0005136.

\bibitem{KMY1}
Y. Kawamura, H. Murayama and M. Yamaguchi, Phys. Lett. B324, 52 (1994).

\bibitem{BIM}
A. Brignole, L.E. Ib\'{a}\~{n}ez and C. Mu\~{n}oz, 
Nucl. Phys. B422, 125 (1994);
Erratum ibid. B437, 747 (1995).


\bibitem{GUT-nonuniversal}
J. Ellis, K. Enqvist, 
D. Nanopoulos, and K. Tamvakis, Phys. Lett. 155B, 381 (1985);
G. Anderson, C.-H. Chen, J.F. Gunion, J. Lykken, T. Moroi and Y. Yamada, 
in New Directions for High Energy Physics, Snowmass 96, 
edited by D. G. Cassel, L. Trindle Gennari, and R. H. Siemann (Stanford Linear
Accelerator Center, Menlo Park, CA, 1997);
K. Huitu, Y. Kawamura, T. Kobayashi, and K. Puolam\"{a}ki, Phys. Rev. D61,035001 (1999);
G. Anderson, H. Baer, C.-H. Chen, P. Quintana and X. Tata, 
Phys. Rev. D61, 095005 (2000).

\bibitem{Hisano-Goto-Murayama}
J. Hisano, T. Goto, and H. Murayama, Phys. Rev. D49, 1446 (1994)

\bibitem{HIY}
T. Hotta, K.-I. Izawa and T. Yanagida, 
Phys. Rev. D53, 3913 (1996); 
Prog. Theor. Phys. 95, 949 (1996); 
Phys. Rev. D54, 6970 (1996). 
 
\bibitem{non-universal-product-GUT}
N. Arkani-Hamed, H.-C. Cheng and T. Moroi, Phys. Lett. B387, 529 (1996);
K. Kurosawa, Y. Nomura and K. Suzuki, Phys. Rev. D60, 117701 (1999).




\bibitem{MizutaNgYamaguchi}
S. Mizuta, D. Ng, and M. Yamaguchi, Phys. Lett. B300, 96 (1993)

\bibitem{1Loopmass} 
D. Pierce and A. Papadopoulos, Nucl. Phys B430, 278 (1994), and 
Phys. Rev. D50, 565 (1994)

\bibitem{ChenDress96}
C.-H. Chen, M. Dress and J. F. Gunion, Phys. Rev. Lett. 76, 2002 (1996);
Phys. Rev. D55, 330 (1997) 330

\bibitem{FengMoroietal99}
J. F. Feng, T. Moroi, L. Randall, M. Strassler, and S. Su, Phys. Rev. Lett. 83, 1731 (1999) 

\bibitem{FengMoroi00}
J. F. Feng and  T. Moroi, Phys. Rev. D61, 095004 (2000) 

\bibitem{GunionMrenna00}
J. F. Gunion and S. Mrenna, Phys. Rev. D62, 015002 (2000) 




\bibitem{MizutaYamaguchi}
S. Mizuta and M. Yamaguchi, Phys. Lett. B298, 120 (1993)

\bibitem{GiudicePomarol96}
G. F. Giudice and A. Pomarol, Phys. Lett. B372, 253 (1996)

\bibitem{DressNojiri97}
M. Dress, M. M. Nojiri, D. P. Roy and Y. Yamada, Phys. Rev. D56, 276 (1997)



\end{references}
\end{document}